\documentclass[final,1p,times,numbers,sort&compress]{elsarticle}
\usepackage{txfonts}

\usepackage{amssymb}
 \usepackage{amsthm}
\usepackage{amscd}
\usepackage{amsmath}
\usepackage{amsfonts}
\usepackage{amssymb}
\usepackage{graphicx}
\usepackage{subfigure}
\usepackage{float}
\usepackage{epsfig}
\usepackage{graphics}
\usepackage{booktabs}
\usepackage{epstopdf}

\numberwithin{equation}{section}

\journal{$\ast\ast\ast$}

\begin{document}

\begin{frontmatter}

\title{Stability of China's Stock Market: Measure and Forecast by Ricci Curvature on Network}

\author[label1]{Xinyu Wang}
\address[label1]{School of Science, China Agricultural University, Beijing, 100091, China}
\author[label2]{Liang Zhao}
 \ead[label2]{ liangzhao@bnu.edu.cn}
\address[label2]{Corresponding author, School of Mathematical Sciences,\\
Key Laboratory of Mathematics and Complex Systems of MOE,\\
Beijing Normal University, Beijing 100875, China}
\author[label3]{Ning Zhang}
\address[label3]{School of Finance, Chinese Fintech Research Center,\\ Central University of Finance and Economics, Beijing, 102206, China}
\author[label3]{Liu Feng}
\author[label1]{Haibo Lin}

\begin{abstract}
The systemic stability of a stock market is one of the core issues in the financial field. The market can be regarded as a complex network whose nodes are stocks connected by edges that signify their correlation strength. Since the market is a strongly nonlinear system, it is difficult to measure the macroscopic stability and depict market fluctuations in time. In this paper, we use a geometric measure derived from discrete Ricci curvature to capture the higher-order nonlinear architecture of financial networks. In order to confirm the effectiveness of our method, we use it to analyze the CSI 300 constituents of China's stock market from 2005--2020 and the systemic stability of the market is quantified through the network's Ricci type curvatures. Furthermore, we use a hybrid model to analyze the curvature time series and predict the future trends of the market accurately. As far as we know, this is the first paper to apply Ricci curvature to forecast the systemic stability of domestic stock market, and our results show that Ricci curvature has good explanatory power for the market stability and can be a good indicator to judge the future risk and volatility of the domestic market.
\end{abstract}


\end{frontmatter}

\section{Introduction}

Through more than thirty years of development, China's capital market has grown continuously. With improvements of the trading mechanism, the market stability has been gradually enhanced and the market plays a more and more important role in optimizing the social financing structure and promoting the allocation of resources. On the other hand, China's financial market is in its infancy, and abnormal market fluctuations still occur occasionally. For example, from 2007 to 2008, the Shanghai Composite Index fell from $6124$, the highest point, to $1664$, a drop of $70\%$. During the market crash in 2015, the market experienced significant abnormal fluctuations which lasted for half a year.  As the key factors of derivative pricing and financial risk management, it is of great significance to study how to measure and forecast the market stability reasonably and accurately. This kind of ability to analyze and predict the market is conducive to the objective and quantifiable evaluation of China's financial market, to the analysis of the market stability factors and the formulation of targeted policies, so as to realize the early warning and prevention of financial risks and the maintenance of financial stability.

The stock market is a nonlinear and non-stationary system with strong volatility, tight coupling and asymmetry. Individual stocks in the market interact each other and the abnormal fluctuations of individuals may quickly enlarge to the whole market. To better understand the highly correlated market, as well as to achieve monitoring and adjustment of it, economists advocate the use of many new tools and interdisciplinary approaches, such as trigger points, feedback, contagion and complexity theory\cite{Suth,Baig,Subr,Bouc,Ghou,Chak}. In particular, to describe the stability macroscopically, we should not consider each individual separately, but should regard the market as a whole system, which coincides with the nature of complex networks\cite{Mant,Batt}. Empirical cross-correlation among stock prices has been extensively studied and explored more than two decades\cite{Pler,Mant1,Lalo,Gopi,Kull,Pler1}. The correlation between stock returns allows us to construct a variety of correlation-based networks, such as minimum spanning trees (MST)\cite{Mant1,Duss,Micc,Onne} or threshold networks\cite{Kuma}, where nodes represent stocks and edges represent correlation strength (or converted to a distance metric). In recent years, correlation-based networks become one of the common tools for modeling and analyzing complex financial systems\cite{Pler1,Onne,Tumm,Phar,Chak1}.

Since there are interactions that occur among groups of more nodes besides pairwise interactions, to reveal the higher-order nonlinear relationship in a network \cite{Krio,Bian,Sree,Sama}, curvature, which is a key concept in geometry proposed by Gauss and Riemann\cite{Jost}, can be an appropriate and powerful tool, and it has been increasingly used as network metrics in recent years\cite{Sree,Sama,Sand,Ni}. In 2015, Sandhu et al.\cite{Sand1} applied the graph curvature to cancer networks for the first time. Sandhu et al.\cite{Sand} also studied the evolution of Ollivier-Ricci curvature in the financial threshold network and showed that Ollivier-Ricci curvature can be used to determine the stability of USA S$\&$P-500 over the period 1998-2013. A recent study by Samal et al.\cite{Sama1} confirms that discrete Ricci curvature can be an excellent indicator of stability and volatility for financial markets of USA and Japan. For the financial market in China, relevant studies have confirmed that it has significant small-world effect and scale-free feature\cite{Jin,Zhang,Wang}, which provides us a theoretical basis for the combination of network geometry and domestic financial market. In summary, the description of the stability of the domestic stock market through geometric measurement is the first motivation of the research work in this paper.

In addition to measure the stability, prediction of trends of the market is also an exciting research area and this is another main purpose of this paper. We will use a hybrid machine learning model combing deep neural network and wavelet decomposition to achieve this goal. We remark that, because the financial curvature time series are complex, non-stationary and very noisy, the classic time series models, such as ARIMA, GARCH, et al., are not suitable for this task.

Since deep learning models can successfully extract features of real-world data, combining deep learning with financial market forecasting is regarded as a charming strategy\cite{Cava}. Among them, recurrent neural network (RNN)\cite{Werb,Hoch} is a kind of recursive neural network that is input from sequence data, recursive in the direction of the evolution of sequence, and chained by all nodes. To overcome gradient disappearance and gradient explosion of RNN, a specific kind of RNN named Long Short--Term Memory (LSTM)\cite{Hoch1,Gers}, which takes into account the long-term dependence of time series, is gradually used in time series forecasting. Kumar and Ningombam\cite{Kuma1} evaluated the effectiveness of LSTM for making predictions about stock prices of APPL(Apple Inc./NASDAQ). Liu\cite{Liu} applied LSTM to the large interval volatility forecasting of S$\&$P 500 and AAPL, and finally concluded that LSTM can achieve a better forecasting result than GARCH(1,1). Huang, et al.\cite{Huan} decomposed financial data into long-term and short-term trends by variational mode decomposition and then utilized LSTM to predict the future trends of the sequences.

Wavelet decomposition (WD) is an approach that describes the relationship between the time series in time and frequency domains simultaneously. Through wavelet decomposition, the noise feature of time series can be fixed. Therefore, it is natural to combine wavelet decomposition and forecasting models to improve the prediction accuracy of time series. In the research of impact of COVID-19 on the global economy, \v{S}tifani\'{c}, et al.\cite{Stif} integrated the stationary wavelet transform and bidirectional long short-term memory neural network to forecast Crude Oil and stock prices and achieved satisfactory results. Peng, et al.\cite{Peng} applied a LSTM-based model into energy consumption forecasting, which also combined wavelet decomposition and LSTM, and achieved better prediction accuracy compared with the basic LSTM model.

In the present paper, according to the above work and our two main purposes, we first construct a threshold network based on the daily returns of the constituents of CSI 300 index over 16 years. A main objective of this study is to confirm that discrete Ricci curvature can be applied to networks of China's stock market and can accurately describe its systemic stability. We find that Ricci curvature provides a good response to the systemic characteristic of the financial market in China and we can use this tool to to identify important events (good or bad) in the market. As another main contribution, we develop a hybrid forecasting model which provides a good response to the future trends of the market.

\baselineskip 15pt
\newpage
\section{Preliminaries}

\subsection{Graph and Minimum Spanning Tree}
In mathematics, we usually call a network a graph, which is composed of a finite set of nodes and a set of edges between nodes, denoted as $G(V, E)$, where $G$ is denoted as a graph, $V$ is the set of nodes in $G$, and $E$ is the set of edges in $G$. Table \ref{concept} list some of the concepts related to graph.

\begin{table}[htp]
\centering
    {\small
    \begin{tabular}{cc}
        \toprule
        \textbf{Professional terminology}   & \textbf{Definition} \\
        \midrule
         Directed Edge     & The edge has directions\\
         Undirected Edge        & The edge has no direction\\
         Directed graph  &  All edges of the graph are directed edges\\
         Undirected graph & All edges in the graph are undirected\\
         Directed complete graph & A directed graph with edges between any two nodes\\
         Undirected complete graph & An undirected graph with edges between any two nodes\\
         Weight & Edge--related numbers\\
        \bottomrule
    \end{tabular}}
    \caption{\small Basic concepts of graphs}
    \label{concept}
\end{table}

For brevity, we only discuss undirected graphs. Two nodes of a graph are said to be connected if there is a path between them. If any two nodes in the graph are connected, the graph is called a connected graph. The spanning tree of a connected graph with $n$ vertices is a connected subgraph that contains all $n$ vertices, but has only $n-1$ edges. If an edge is added to a spanning tree, it necessarily forms a ring, and if an edge is reduced, it is no longer a connected graph.

{\bf Minimum Spanning Tree(MST)}: In a given undirected graph $G = (V, E)$, $e_{uv}$ represents the edge connecting nodes $u$ and $v$, and $\omega_{uv}$ represents the weight of this edge. If there exists $T$ which is a spanning tree of $G$ and $\omega(T)$ is minimal, $T$ is called a minimal spanning tree of $G$. We usually use Prim's algorithm\cite{Prim} to implement the construction of minimum spanning trees of a graph.

\subsection{Ricci-type Curvatures for Network Analysis}
As an important geometric quantity, the classical Ricci curvature quantifies the deviation for the tangent direction and requires a smooth manifold as well as a tensor and higher order derivatives\cite{Jost}. This requirement is not applicable to discrete graphs or networks, so it is necessary to discretize it to apply in networks. In this work, we apply four different types of discrete Ricci curvatures to the threshold network of China's stock market. Their definitions and applications can be found in many relevant literatures. For completeness, we briefly describe their definitions here.

{\bf Ollivier-Ricci Curvature}:
This is a widely used discretization\cite{Sama,Ni,Sand} of the classical Ricci curvature raised by Olliver\cite{Olli,Olli1}. In recent years it has also been applied to financial networks\cite{Sand1,Sama1}. In a space with positive curvature, the average distance between balls is less than the center distance, while in a negative curved space, the opposite conclusion is reached. Ollivier-Ricci(OR) curvature extends the above observations from balls (volumes) to measures (probabilities), and the OR curvature of the edge $e$ connecting nodes $u$ and $v$ is defined as
\begin{eqnarray}\label{OR}
O(e)=1-\frac{W_1(m_u,m_v)}{d(u,v)}.
\end{eqnarray}
In \eqref{OR}, $m_u$ and $m_v$ represent measures concentrated at nodes $u$ and $v$, $d(u,v)$ is the distance between $u$ and $v$, and $W_1$ is the Wasserstein distance\cite{Vase} between the discrete probability measures $m_u$ and $m_v$. The Wasserstein distance is given by
\begin{eqnarray*}\label{Wasserstein}
W_1(m_u,m_v)=\underset{\mu_{u,v} \in \Pi (m_u, m_v) }{\inf} \sum_{(u',v')\in V \times V}d(u',v')\mu_{u,v}(u',v'),
\end{eqnarray*}
where $\Pi(m_u, m_v)$ is the set of probability measures $\mu_{u,v}$ that satisfy
\begin{eqnarray*}
\sum_{u' \in V}\mu_{u,v}(u',v')=m_v(v'),~~~\sum_{v' \in V}\mu_{u,v}(u',v')=m_u(u')
\end{eqnarray*}
In addition, the probability distribution $m_u$ for $u \in V$ must be specified, which is chosen to be uniform over the neighboring nodes of $u$\cite{Lin}.

{\bf Forman-Ricci Curvature}:
Forman-Ricci(FR) Curvature is based on the relationship between the Riemannian Laplace operator and the Ricci curvature\cite{Form}. It has been shown that FR curvature and edge betweenness centrality are highly correlated\cite{Sama,Sree1}. In the undirected network, the FR curvature of edge $e$ connecting nodes $u$ and $v$ is defined as\cite{Sree}
\begin{eqnarray}\label{FR}
F(e)=\omega_e \left( \frac{\omega_{u}}{\omega_e}+\frac{\omega_{v}}{\omega_e}- \sum_{e_{u}\sim e,e_{v}\sim e}\left[ \frac{\omega_{u}}{\sqrt{\omega_{e}\omega_{e_{u}}}}+\frac{\omega_{v}}
{\sqrt{\omega_{e}\omega_{e_{v}}}}\right]\right),
\end{eqnarray}
where $\omega_e$, $\omega_{u}$ and $\omega_{v}$ denote the weights of the edge $e$, the nodes $u$ and $v$ respectively. In addition, $e_{u} \sim  e$ and $e_{v} \sim e$ denote the set of edges connecting $u$ and $v$, respectively, but excluding the edge $e$.

{\bf Menger-Ricci Curvature}:
Menger's approach\cite{Meng} is based on viewing the graph as a metric space, and the path length between two nodes is treated as the distance between two points in the metric space. Suppose $T$ is a triangle in the metric space with sides $a$, $b$ and $c$, then Menger curvature of $T$ is given by
\begin{eqnarray*}\label{Me}
M(T)=\frac{1}{R(T)}=\frac{\sqrt {p(p-a)(p-b)(p-c)}}{a \cdot b \cdot c},
\end{eqnarray*}
where $p = (a + b + c)/2$ and $R(T)$ is the radius of the circumscribed circle of the triangle $T$. Then, Menger-Ricci(MR)
curvature of an edge $e$ in a network can be defined as\cite{Sauc}
\begin{eqnarray}\label{MR}
M(e)=\sum_{T_e \sim e} M(T_e),
\end{eqnarray}
where $T_e \sim e$ denotes the set of triangles formed by side $e$.

{\bf Haantjes-Ricci Curvature}:
Haantjes\cite{Haan} defined the curvature of a curve in a metric space as the ratio of the arc length to the chord length of the curve. For a discrete network, suppose that $\pi=v_0, v_1, \cdots v_n$ is a simple path between nodes $v_0$ and $v_n$, $l(\pi)$ is the length of the path and $d(v_0, v_n)$ is the shortest distance between nodes $v_0$ and $v_n$. Haantjes-Ricci(HR) curvature of the simple path $\pi$ is
\begin{eqnarray*}\label{formula8}
H^2(\pi)=\frac{l(\pi)-d(v_0, v_n)}{d(v_0,v_n)^3}.
\end{eqnarray*}
Then, HR curvature of an edge $e$ can be defined as
\begin{eqnarray}\label{HR}
H(e)=\sum_{\pi \sim e} H(\pi),
\end{eqnarray}
where $\pi \sim e$ denote the paths that connect the nodes anchoring the edge $e$.

The above four discretizations focus on capturing different geometric properties portrayed by the classical Ricci curvature. OR curvature can well capture the aspect of volume growth of classical Ricci curvature. We use OR curvature in networks to compare the average distance between two nodes. FR curvature depicts the geodesic diffusivity of the classical Ricci curvature and we use FR curvature in networks to show the information spread at the ends of edges. Both MR and HR curvatures can capture the geodesics dispersal rate of the classical Ricci curvature. In this work, we ignore the weights of the edges in the network and calculate the average of edges for these four discrete Ricci curvatures according to equations (\ref{OR}-\ref{HR}), respectively, and considering the computational complexity, we only use the path between nodes whose length is less than or equal to $4$ in the calculations of MR and HR curvatures.

\subsection{Discrete Wavelet}
Wavelet analysis is a time-frequency analysis method and can achieve high resolution in both time and frequency domains. Through decomposing the curvature time series of our financial networks into several components based on various frequencies, wavelet analysis is able to filter out the chaotic components, so as to remove the influence of noises and improve the prediction performance effectively.

The wavelet transform is roughly divided into continuous transform and discrete transform and both are based on two specific functions: mother wavelet function and daughter wavelet function. For the continuous case, assuming $\psi \in L^2(\mathbb{R})$ and $\widetilde{\psi}(\omega)$ is the Fourier transform of $\psi(t)$, $\psi(t)$ is called mother wavelet function, if $\widetilde{\psi}(\omega)$ meets:
\begin{eqnarray*}
C_{\psi}=\int \frac{|\widetilde{\psi}(\omega)|^2}{|\omega|}d \omega < \infty.
\end{eqnarray*}
And the definition of daughter wavelet function is as followed:
\begin{eqnarray}\label{cwf}
\psi_{a,b}(t)=\frac{1}{\sqrt{|a|}}\psi \left( \frac{t-b}{a} \right),
\end{eqnarray}
where $a$ and $b$ are respectively called expansion factor and translation factor.

Due to the fact that our curvature data is based on the daily returns of stocks, we utilize the discrete wavelet transform to decompose the time series. Assigning $2^{-j}$ and $k2^{-j}$ to $a$ and $b$ in equation\eqref{cwf}, discrete daughter wavelet function is as followed:
\begin{eqnarray}\label{dfunction}
\psi_{2^{-j},k2^{-j}}(t)=2^{j/2} \psi(2^{j}t-k),
\end{eqnarray}
where $j, k \in \mathbb{Z}$. For brevity, we use $\psi_{j,k}(t)$ instead of $\psi_{2^{-j},k2^{-j}}(t)$ from now on. The discrete wavelet transform corresponding $\psi_{j,k}(t)$ is as followed:
\begin{eqnarray}\label{dwf}
DWf(j,k)=\langle f, \psi_{j,k} \rangle= 2^{j/2} \int_{-\infty}^{+\infty} f(t) \overline{\psi}(2^{j}t-k)dt,
\end{eqnarray}
where $f(t) \in L^2(\mathbb{R})$ and $\overline{\psi}$ is the conjugate of $\psi$.

Our denoising process of wavelet decomposition is divided in to the following three steps:

{\bf Step1}: determine a wavelet function and the number of decomposition layers, and then decompose the original time series.

{\bf Step2}: select an appropriate threshold to eliminate the fluctuation exceeding the threshold and retain the specific signals.

{\bf Step3}: reconstruct the retained signals to form a new signal.

\section{Data and Methods}
\subsection{Data Description}
The data of this paper are collected from Eastmoney(www.eastmoney.com), including daily closing prices for $N=111$ stocks, $T=3889$ trading days, from January 4, 2005 to December 31, 2020. All the $N=111$ stocks are constituents of CSI 300 Index. Due to some unavoidable factors such as stock suspensions, some stocks are missing their prices on certain trading days. Considering that the stock prices do not change too much in a short period of time, we fill the gaps with the data of previous trading time.

First for each stock, we construct a daily return time series $r_k(t)$ according to the formula as followed:
\begin{eqnarray*}
r_k(t)=\ln P_k(t)-\ln P_k(t-1),
\end{eqnarray*}
where $k=1,2,\cdots, N$, $t=2,3,\cdots, T$ and $P_k(t)$ is the adjusted closing price of the $k$th stock at time $t$. Then, the equal-time Pearson cross-correlation coefficients $c_{ij}$ of the daily return time series of stock $i$ and stock $j$ is defined as
\begin{eqnarray*}
c_{ij}(t)=\frac{Cov(r_i, r_j)}{\sigma_i \sigma_j},
\end{eqnarray*}
where $Cov(r_i,r_j)$ is the covariance of $r_i$ and $r_j$ in a time interval of length $\tau$, $i, j = 1, \cdots, N$, $t$ indicates the end date of the interval of $\tau$ trading days. In our empirical research, we use the following two schemes to divide time series in order to better illustrated the reliability of our conclusion by comparison of these two approaches.

(i) A non-overlapping time interval of $\tau$=22 trading days (one trading month),

(ii)An overlapping time interval of $\tau$=22 days, with a rolling shift of $\Delta$=5 trading days (one trading week).

Corresponding to correlation coefficients, we construct the distance measures $d_{ij}$ which are widely used for the construction of financial networks\cite{Mant1,Mant2,Onne}.
\begin{eqnarray*}
d_{ij}(t)=\sqrt{2(1-c_{ij}(t))}.
\end{eqnarray*}

\subsection{Threshold Network Construction}
Firstly, for a given time interval of $\tau$ trading days ending on trading day $t$, we get a distance matrix $D_{\tau}(t)$ whose elements are $d_{ij}(t)$.  This distance matrix $D_{\tau}(t)$ can be considered as an edge-weighted complete graph $G_{\tau}(t)$, whose nodes are stocks and the weight of an edge between stocks $i$ and $j$ is given by $d_{ij}(t)$. Next, with the help of Prim's algorithm\cite{Prim}, we create MST $T_{\tau}(t)$ based on the complete graph $G_{\tau}(t)$, which selects the most relevant connections of the stocks. Finally, to capture more significant information in the market, we add edges in $G_{\tau}(t)$ to connect corresponding nodes $i$ and $j$ in $T_{\tau}(t)$ if $c_{ij}(t) > \theta$ for some threshold $\theta$. The complete graph constructed by MST and the threshold $\theta$ is called threshold network and is denoted as $S_\tau (t)$.

In this paper, we set the threshold $\theta = 0.75$ and use $S_{\tau}(t)$ for calculating different kinds of Ricci curvatures.

\subsection{The Hybrid Forecasting Model}

\begin{figure}[htp]
  \centering
  {\includegraphics[width=7.5cm]{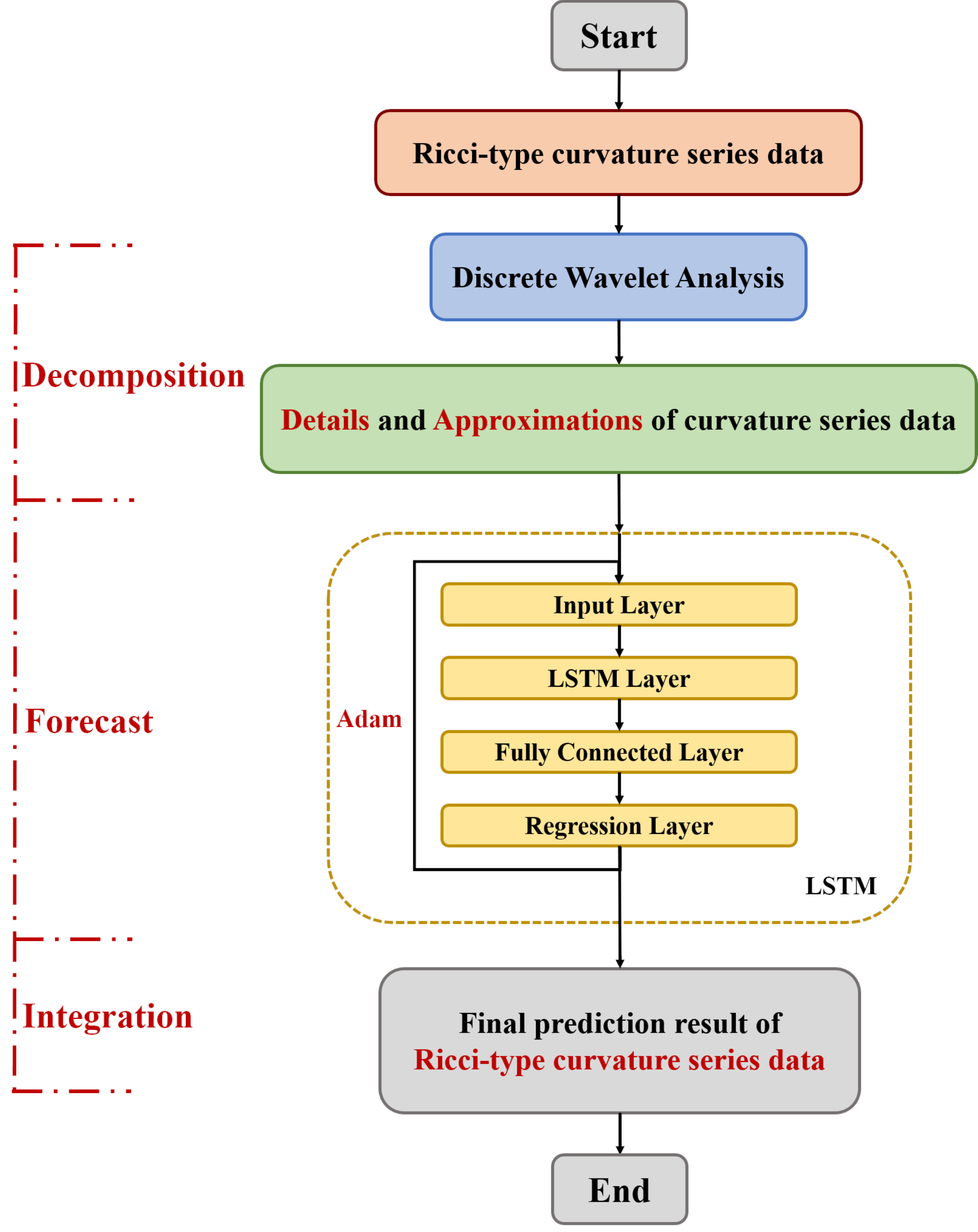}}
  \caption{\small The architechture of WD-LSTM}
  \label{f-walstm}
\end{figure}

Due to the fact that the curvature time series is composed of nonlinear features, various temporal information and noises, it is challenging to achieve an accurate forecasting result. Wavelet decomposition can analyze the series from different scales, which can not only reflect the overall trend, but also extract the effective information of the series in details. On the other hand, as a deep learning model, LSTM is able to learn long-term correlations and mine complicated nonlinear relationships within the curvature series effectively. Based on the above facts, we propose a hybrid WD-LSTM model, combining the strengths of wavelet decomposition and long short-term memory network, to forecast the future trends of the market. The WD-LSTM model involves three phrases: decomposition, forecasting and integration. In the decomposition phrase, we decompose the original curvature series data into four high frequency sequences (detail) and one low frequency sequence (approximation). Next, in the forecasting phrase, LSTM is utilized to forecast each decomposed sequence respectively. Finally, the prediction results of all sub-sequences are aggregated in the integration phrase. The architecture of the WD-LSTM model is shown in Figure \ref{f-walstm}.

\begin{figure}[htp]
  \centering
  {\includegraphics[width=12cm]{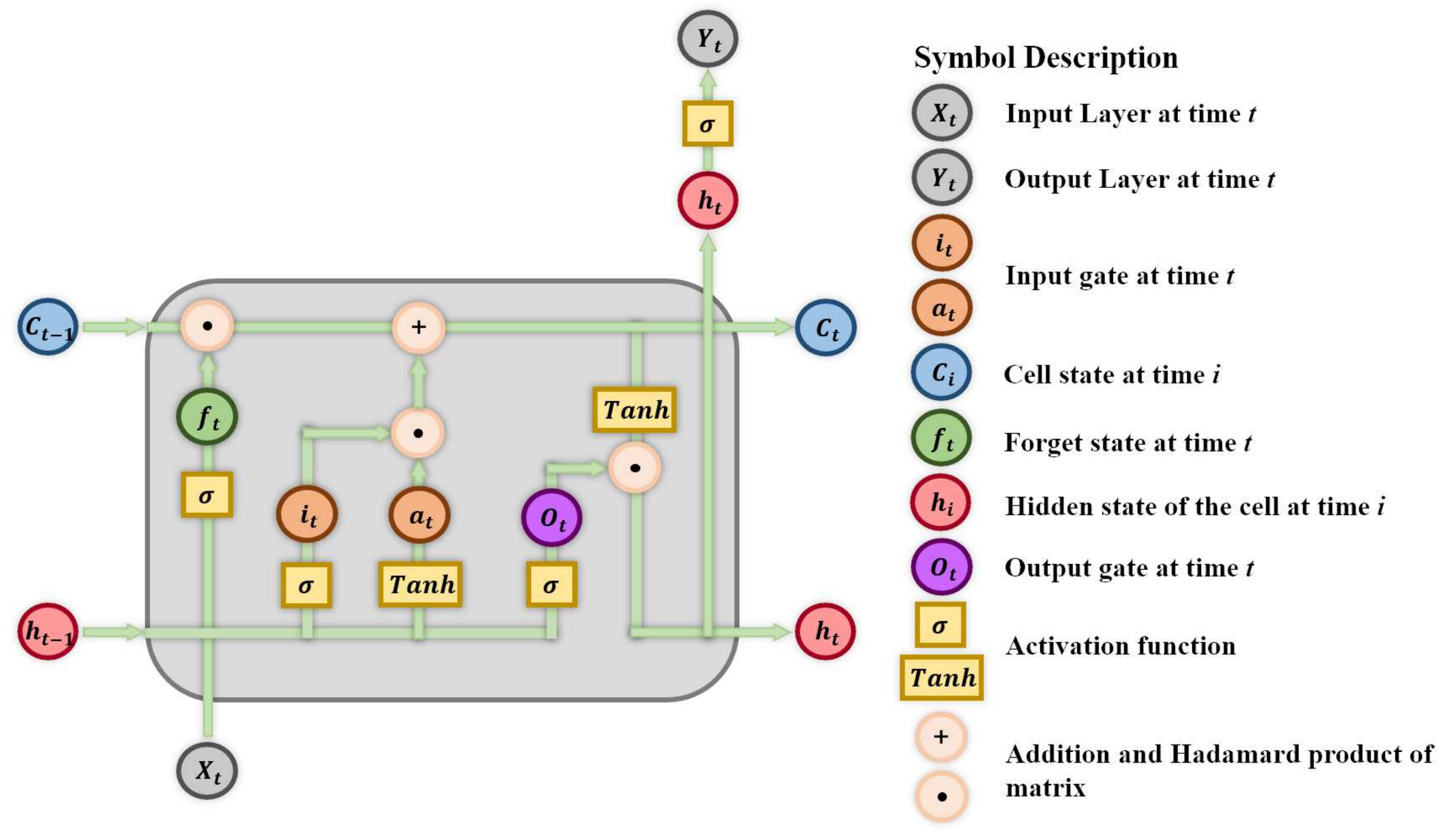}}
  \caption{\small The architechture of LSTM}
  \label{f-lstm}
\end{figure}

LSTM used in the forecasting phrase is a specially designed RNN and suitable for processing and forecasting important events with very long intervals and delays in the time series. The architecture of LSTM at time $t$ is composed of four units: forget gate, input gate, output gate and cell state, which is shown in Figure \ref{f-lstm}. To clarify the details of LSTM, we use $W$, $U$ and $b$ with different subscripts to denote the linear coefficients and biases of these units.

The output $f_t$ of forget gate at time $t$ represents the probability of forgetting the hidden cell state of the previous layer, which can be calculated by:
\begin{eqnarray*}
f_t=(\sigma W_f h_{t-1}+U_f x_t +b_f),
\end{eqnarray*}
where $\sigma$ is the sigmoid activation function, $h_{t-1}$ denotes the state of hidden layer at time $t-1$, $x_t$ denotes the input vector at time $t$.

The input gate is responsible for processing the current input signal and composed of two parts depending on sigmoid and $\tanh$ activation functions respectively. This gate can be formulated as:
\begin{eqnarray*}\label{formula18}
\left\{
\begin{aligned}
i_t&=\sigma (W_i h_{t-1}+U_i x_t +b_i)\\
a_t&=\tanh(W_a h_{t-1}+U_a x_t+b_a)\\
\end{aligned}
\right.
\end{eqnarray*}

The cell state is updated according to forget gate and input gate which is formulated as:
\begin{eqnarray*}\label{formula19}
C_t=C_{t-1} \odot f_t + a_t \odot i_t,
\end{eqnarray*}
where $\odot$ denotes the Hadamard product.

The output gate is formulated as:
\begin{eqnarray*}
\begin{aligned}
O_t&=\sigma (W_o h_{t-1}+U_o x_t +b_o)\\
\end{aligned}
.
\end{eqnarray*}
With the output state $O_t$ and hidden cell state $C_t$ at time $t$, the hidden state of the cell is updated as:
\begin{eqnarray*}
\begin{aligned}
h_t&=O_t \odot \tanh(C_t)\\
\end{aligned}
.
\end{eqnarray*}

Finally, we set a forecast unit which is a fully connected neural network with outputs be the forecasting values $Y_t$ of the time series at time $t+1$ according to the hidden state $h_t$:
\begin{eqnarray*}\label{formula21}
Y_t= \sigma (Wh_t+ b).
\end{eqnarray*}

To complete the building of the whole LSTM model, we set four layers including input layer, LSTM layer, fully connected layer and regression layer, as shown in Figure \ref{f-walstm}, where the regression layer is used to give the mean square error of the outputs.

\section{Empirical Results}
\subsection{Market Stability}
\begin{figure}[htp]
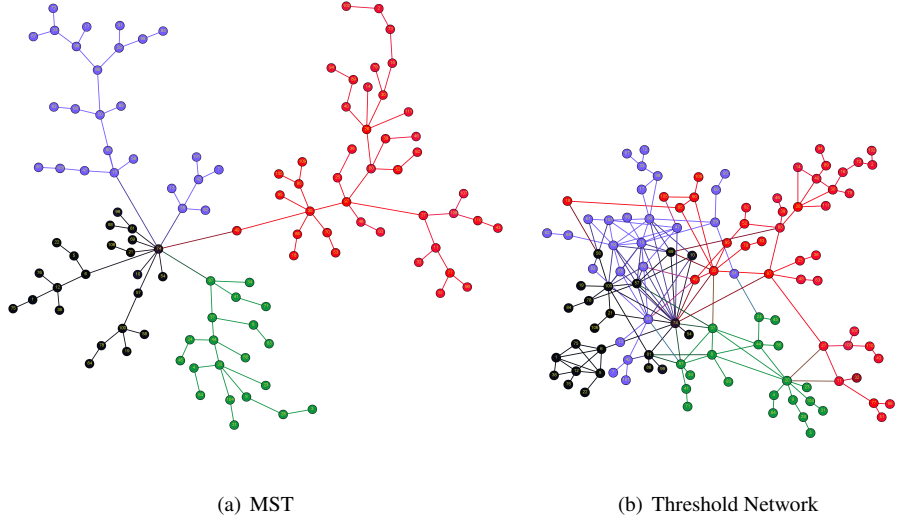

  \centering
  \subfigure[MST]{\includegraphics[width=7cm]{MST.pdf}}
  \subfigure[Threshold Network]{\includegraphics[width=5cm]{Threshold.pdf}}
  \caption{\small The MST and threshold network}
  \label{graph}
\end{figure}

Exploring the explanatory power of Ricci curvature for the stability of China's stock market is one of the main purposes of this paper. We analyze the logarithmic returns of constituents of CSI 300 index over a 16-year period (2005--2020) by means of building the undirected network $S_{\tau}(t)$ with the threshold $\theta=0.75$. The MST and threshold network constructed based on the data is shown in Figure \ref{graph} and Tabel \ref{name} lists some of the ticker symbols corresponding to numbers of nodes in the figure.

\begin{table}[htp]
\centering
    {\small
    \begin{tabular}{ccccccccccc}
        \toprule
        \textbf{Number}   & 18 & 27 & 35 & 49 & 58 & 74 & 97 & 108 \\
        \midrule
        \textbf{Ticher Symbol}   & 600109 & 600183 & 600346 & 600570 & 600703 & 601607 & 000786 & 002008 \\
        \bottomrule
    \end{tabular}}
    \caption{\small List of some of the ticker symbols}
    \label{name}
\end{table}

\begin{figure}[htp]
  \centering
  {\includegraphics[width=11cm]{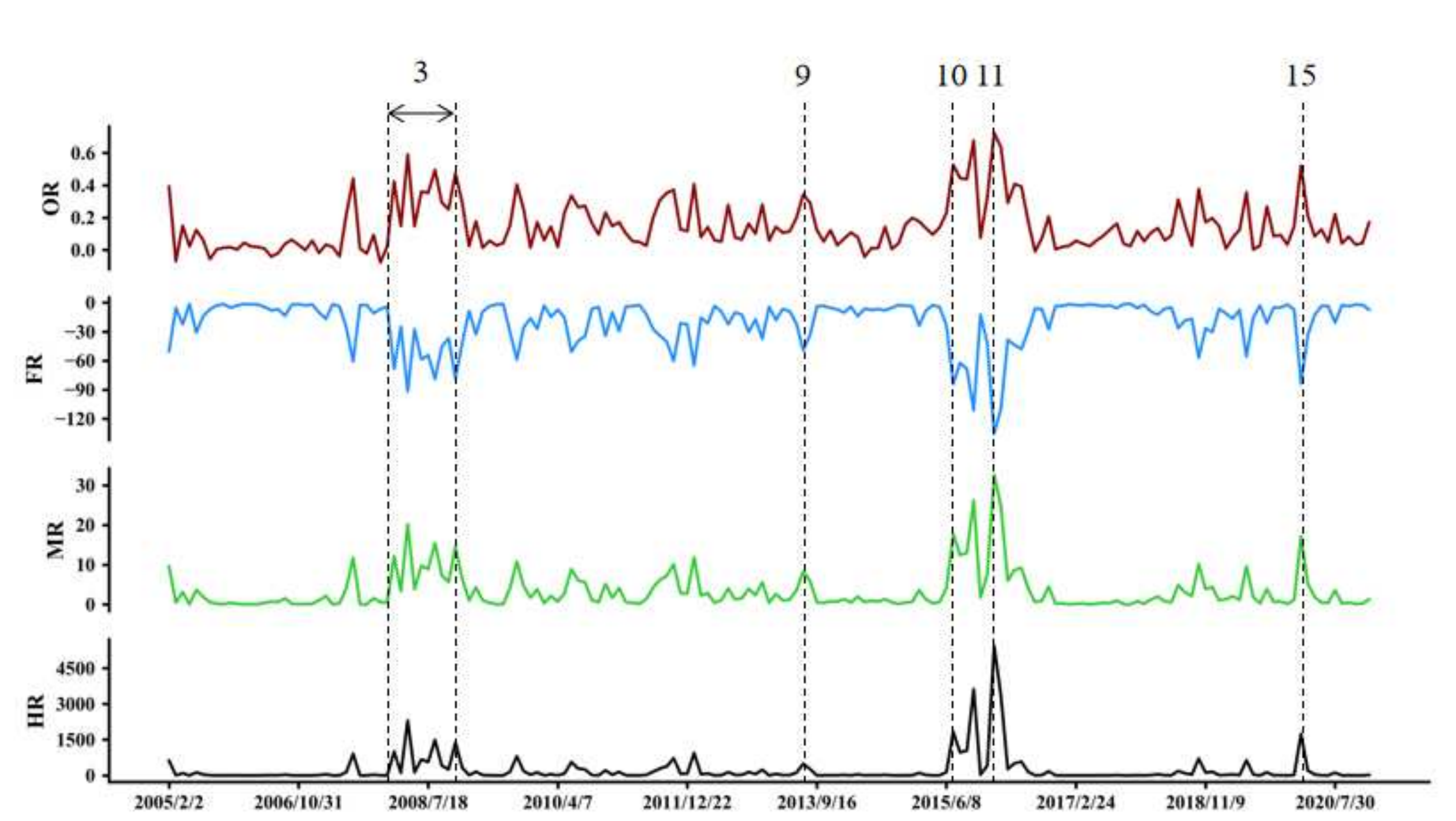}}
  \caption{\small Type (i) curvature time series}
  \label{type1}
\end{figure}

\begin{figure}[htp]
  \centering
  {\includegraphics[width=11cm]{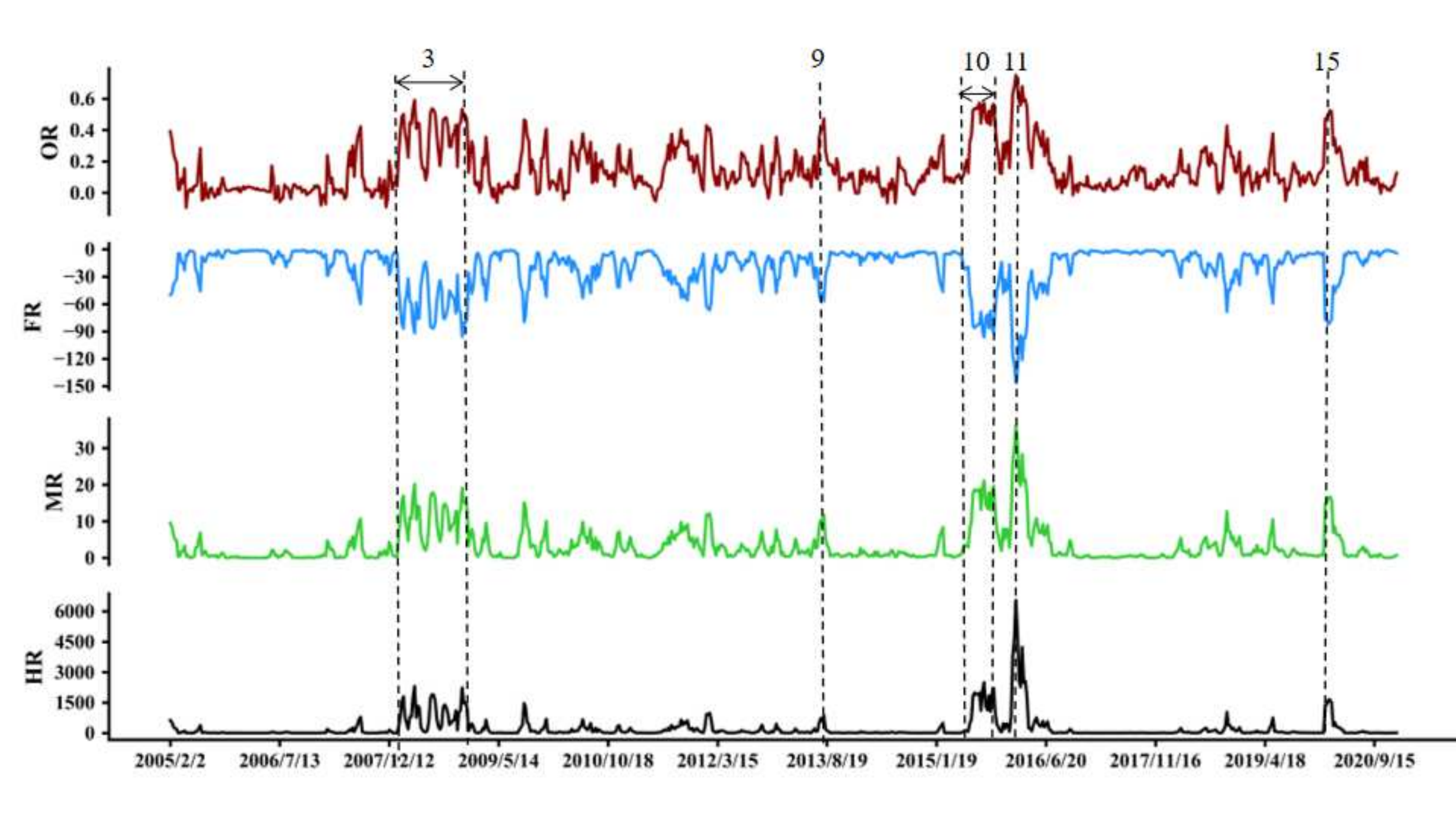}}
  \caption{\small Type (ii) curvature time series}
  \label{type2}
\end{figure}

Figure \ref{type1} depicts four curvature time series of the threshold network $S_{\tau}(t)$ building with non-overlapping time intervals ($\tau=22$ trading days) and Figure \ref{type2} with a rolling shift of $\Delta=5$. Obviously, the fluctuation trends of the curvature time series which are obtained by using two different data processing methods are essentially consistent, which confirms the generalization performance of our methods and the reliability of our conclusions.

We list some of the major events in China's financial market between 2005 and 2020 in Table \ref{goodbad}. As key events in the market, during these events, the rule, structure, participants or external environment of the market have changed significantly and the stability should be poorer than the normal periods. To verify the effectiveness of the geometric quantities of networks, we compare these events and the curvature time series, and find out that the fluctuations of the curvature time series can capture these key information of the market well. Some of the events are marked with dotted lines in Figure \ref{type1} and \ref{type2}.

\begin{table}[htp]
\centering
    {\small
    \begin{tabular}{ccc}
        \toprule
        \textbf{Number}   & \textbf{Events} & \textbf{Time/Period} \\
        \midrule
        1&Shareholding Reform & May 2005\\
        2&Subprime mortgage crisis & Aug 2007\\
        3&International Financial Crisis & 2008-2009\\
        4&Establishment of GEM & 30 Oct 2009\\
        5&First CSI 300 futures contracts listed&16 Apr 2010\\
        6&CSRC proposed eight key tasks&14 Jan 2011\\
        7&PBOC cut RMB RRR&30 Nov 2011\\
        8&Suspension of IPO&2013\\
        9&The mix-up event of Everbright Securities&16 Aug 2013\\
        10&Market Crash in China&15 Jun-9 Jul 2015\\
        11&Implementation of the meltdown mechanism&1 Jan 2016\\
        12&Establishment of the STAR Market &5 Nov 2018\\
        13&Launch of Shanghai-London Stock Exchange&17 Jun 2019\\
        14&First listing of the STAR stocks&22 Jul 2019\\
        15&Impact of COVID-19&3 Mar-1 May 2020\\
        \bottomrule
    \end{tabular}}
    \caption{\small List of some market events between 2005 and 2020}
    \label{goodbad}
\end{table}

Combining the results in Figure \ref{type1} and \ref{type2}, and the events in Table \ref{goodbad}, we find that the four discrete Ricci curvatures can depict the market stability. During the periods of those key events, the curvature time series fluctuates to different degrees. In particular, when the news is significantly good or bad, the time series shows large fluctuations. We therefore believe that the discrete Ricci curvatures can serve as good indicators of the stability for China's stock market.

\subsection{Forecasting of the Systemic Stability}
To accomplish another main purpose, we apply the WD-LSTM model to analyze the curvature time series and forecast the future trend of China's stock market. The WD-LSTM model contains three phrases: decomposition, forecast and integration. The empirical results through the above three phrases are presented below in details.

\begin{figure}[htp]
  \centering
  {\includegraphics[width=11cm]{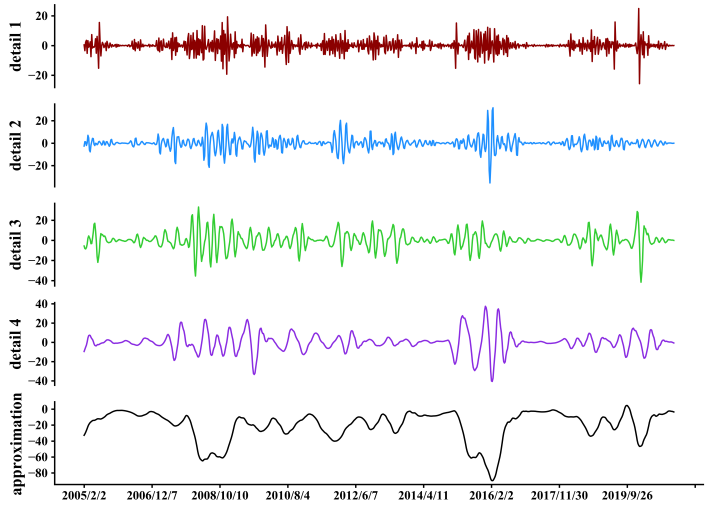}}
  \caption{\small The decomposition result of FR curvature series}
  \label{dec}
\end{figure}

{\bf Decomposition of Curvature Series}:
According to \eqref{dfunction} and \eqref{dwf}, we first decompose the original curvature series into four high frequency sequences (detail) and one low frequency sequence (approximation). For brevity, we choose FR curvature series ($\Delta=5$) as an example and present its decomposition results in Figure \ref{dec}.

{\bf Forecast of Decomposed Sequences}:
The second step of the WD-LSTM model is to forecast each component decomposed by the WD module by using the LSTM module. In our experiment, each decomposed sequences is divided into training set and testing set according to the proportion of $80\%$ and $20\%$. Since $\tau=22$ and $\Delta=5$, the training time series is from February 2, 2015 to November 2, 2017. The number of LSTM layer is set to be $200$. While in the process of training the LSTM model, the max iteration and the initial learning rate is set to be $250$ and $0.005$. Besides, the optimizer of LSTM is chosen to be Adam and the gradient threshold is set to be $1$. After training by using the back-propagation algorithm, we use the hidden state $h_{t-1}$ to forecast the value at time $t$, where $t$ is from November 9, 2017 to December 31, 2020.

\begin{figure}[htp]
  \centering
  {\includegraphics[width=11cm]{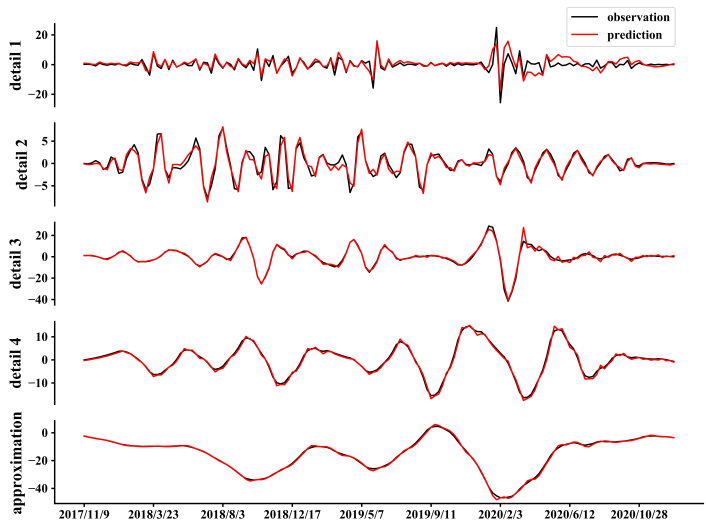}}
  \caption{\small The forecasting result of decomposed sequences (FR curvature)}
  \label{for}
\end{figure}

Figure \ref{for} presents the forecasting result of decomposed sequences of FR curvature series.

\begin{figure}[htp]
  \centering
  {\includegraphics[width=12cm]{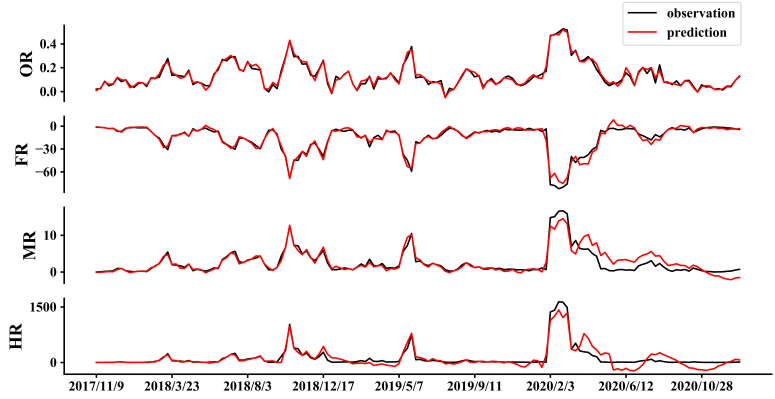}}
  \caption{\small The final forecasting results of four curvature series}
  \label{final}
\end{figure}

{\bf Integration of Forecasting Results}:
The final step of the WD-LSTM model is to integrate the forecasting results of decomposed sequences. After the integration phrase, we can get the final forecasting results of the curvature series. We show the final forecasting results of the four Ricci-type curvature series in Figure \ref{final}. We also list the evaluation metrics of the final forecasting results, including mean absolute error (MAE), mean square error (MSE) and $R^2$, in Table \ref{evaluation}.

\begin{table}[htp]
\centering
    {\small
    \begin{tabular}{ccccc}
        \toprule
        &\textbf{OR}   & \textbf{MR} & \textbf{HR}&\textbf{FR} \\
        MAE& 0.0156& 0.5409& 73.9460& 2.4413\\
        MSE& 0.0004& 0.8745& 14087.5311&18.1357\\
        $R^2$& 0.9653& 0.9295& 0.8701& 0.9459\\
        \bottomrule
    \end{tabular}}
    \caption{\small The evaluation metrics of the WD-LSTM model}
    \label{evaluation}
\end{table}

\subsection{Model Comparison $\&$ Empirical Summary}

To verify the superiority of the WD-LSTM model, in this subsection, we carry out a comparative experiment where a basic LSTM model is utilized to forecast the four Ricci-type curvature series directly. Table \ref{single} presents the evaluation metrics of the single LSTM model's final forecasting results.

\begin{table}[htp]
\centering
    {\small
    \begin{tabular}{ccccc}
        \toprule
        &\textbf{OR}   & \textbf{MR} & \textbf{HR}&\textbf{FR} \\
        MAE&  0.0950& 2.3464& 200.4158& 9.8831\\
        MSE& 0.0146& 14.2501& 186815.2867& 251.1875\\
        $R^2$& -0.1271& -0.1495& -0.7230& 0.2509\\
        \bottomrule
    \end{tabular}}
    \caption{\small The evaluation metrics of the single LSTM}
    \label{single}
\end{table}

Comparing Table \ref{evaluation} and Table \ref{single}, it is obvious that for each evaluation metric, the forecasting performance of the WD-LSTM model is significantly better than that of the basic LSTM model for all the four Ricci-type curvature series. It implies that the wavelet decomposition plays a remarkable role and the hybrid model can handle the strong nonlinearity, complex time characteristics and noise interference of the curvature series better than the single LSTM model.

Furthermore, there must be performance differences between the four Ricci curvatures. Samal, et al.\cite{Sama1} have shown that FR curvature are more sensitive and can detect both crashes and bubbles in USA S$\&$P-500 and Japanese Nikkei-225 markets more efficiently. For China's stock market, comparison of $R^2$ metrics for the four kinds of curvatures in Table~\ref{evaluation} and Table~\ref{single} obviously implies that the performance of the hybrid model is better than a single LSTM model. In the hybrid model, all the four curvatures have excellent explanatory power for depict and forecast the stability of China's stock market. In particular, the $R^2$ metric of OR curvature series is closer to $1$ than those of the other three. We can infer that the OR curvature series is more suitable for the domestic market, which is different from the conclusion about the foreign market. This may reflect the different characteristics of domestic and foreign markets. According to the definitions of these two curvatures, FR curvature is mainly aimed at capturing the diffusion characteristics of the geodesic, which is more sensitive to events than other curvatures, and can better capture the details of the market. While OR curvature measures the relative distance between two respective neighborhoods of two vertices that form an edge. Therefore, it is more suitable for the domestic market where macro-control measures are implemented more effectively and the co-movement effect of the stock sectors is more obvious.

\section{Conclusion}
In this paper, we apply different types of discrete Ricci curvatures of networks to characterize the systemic stability of China's stock market. We verify the reliability of our methods by monitoring the fluctuations of the constituents of CSI 300 index from 2005 to 2020 in conjunction with Table \ref{goodbad}. We find that network curvatures can be used as good indicators for the systemic stability of China's stock market.

Based on the above, we also make a more in depth application of the geometric measure. A hybrid WD-LSTM model, combing wavelet decomposition with long short-term memory network, is applied to forecast the future trends of the systemic stability for China's stock market by means of modeling and predicting the curvature series data. Comparing to the single LSTM model, the WD-LSTM model performs significantly better. Moreover, the empirical result shows that OR curvature is most suitable for the domestic market and the proposed hybrid model has excellent forecasting performance.

In summary, we use discrete Ricci curvature as a measure of the stability for China's financial market and apply an effective hybrid model to forecast the future trends. Our methods and models are very helpful to develop new financial regulatory tools to better identify, forecast, and prevent market risks and contribute to financial stability.

{\bf Acknowledgements.} This research was supported by New Liberal Arts Research and Reform Practice Project of Ministry of Education (NO. 2021060011) and the Emerging Interdisciplinary Project of CUFE.

\end{document}